\documentclass[showpacs,twocolumn]{revtex4} 

\usepackage{graphicx}

\usepackage{dcolumn}
\usepackage{bm}

\begin{document}

\title{Multiphoton Femtosecond Coherent Control in the Single-Cycle Regime}


\author{Lev Chuntonov, Avner Fleischer, and Zohar Amitay}
\email{amitayz@tx.technion.ac.il} %
\affiliation{Schulich Faculty of Chemistry, Technion - Israel
Institute of Technology, Haifa 32000, Israel}

\begin{abstract}
Coherent control of the atomic two-photon absorption with shaped single-cycle pulses is examined theoretically in the weak-field regime. The
control over the stabilized carrier-envelope phase (CEP) of the pulses is determined as a key parameter allowing the full utilization of the
ultrabroad pulse spectrum. This bandwidth is sufficient to provide besides the sequential two-photon pathways, additional Raman-type pathways
coupling between the ground and the excited states. The interference between the different pathways is efficiently controlled via the control
over the CEP. Simplified two-level model is analyzed in the frequency domain where the rational pulse shaping is applied. The developed
intuition is applied to atomic Cs and verified by the exact numerical solution of time-dependent Schr\"{o}dinger equation.
\end{abstract}


\maketitle

Ultrashort laser pulses are a unique experimental tool to explore
and control the ultrafast dynamics of the quantum systems
\cite{tannor_kosloff_rice_coh_cont, shapiro_brumer_coh_cont_book,
dantus_exp_review1_2, silberberg_2ph_nonres1_2, baumert_2ph_nonres,
silberberg_2ph_1plus1, girard_2ph_1plus1,
silberberg_antiStokes_Raman_spect, leone_res_nonres_raman_control,
gersh_murnane_kapteyn_Raman_spect, amitay_2ph_inter_field1}.
The ultrashort duration of the pulses is originated from their
ultrabroad coherent spectral bandwidth which provides a manifold of
photoinduced multiphoton pathways between the ground and the excited
states of the system.
The relative phase, amplitude, and polarization of different pulse
spectral components are used to manipulate the interference between
the pathways.
These characteristics are assigned to the spectral components using
pulse shaping techniques \cite{pulse_shaping} and act as control
parameters for enhancing or attenuation of the transition
probability to the excited state.
In particular, shaped femtosecond pulses are extensively used for
experimental coherent control of bound-bound multiphoton transitions
in atoms and molecules at different regimes of the excitation pulse
intensities \cite{dantus_exp_review1_2, silberberg_2ph_nonres1_2,
baumert_2ph_nonres, silberberg_2ph_1plus1, girard_2ph_1plus1,
silberberg_antiStokes_Raman_spect, leone_res_nonres_raman_control,
gersh_murnane_kapteyn_Raman_spect, amitay_2ph_inter_field1}.
Recently, the pulses as short as few optical cycles became widely
available \cite{krause_single_cycle_1, yamashita_single_cycle,
akturk_single_cycle, rausch_single_cycle, krause_single_cycle_2}.
One of the basic characteristics of the few-cycle pulses is the
relative phase between the spectral components of the pulse and its
spectral envelope.
In the time-domain, this global phase is translated to the relative
phase between the carrier wave of the pulse and its temporal
envelope.
Carrier-envelope phase (CEP) plays significant role in the many
multiphoton processes like photoionization in the various
pulse-intensity regimes, high harmonics generation, and others
\cite{krause_single_cycle_2, nakajima_phase_dependent,
wu_phase_dependent, chelkowski_phase_dependent,
krause_phase_dependent, bucksbaum_phase_dependent_2}.
It is also a very important parameter to be considered for the
coherent control of bound-bound multiphoton transitions with the
single-cycle pulses.
In the present work we demonstrate the role of the CEP on the
multiphoton absorption in the weak-field limit.
We show that CEP stabilization is necessary to obtain full control
over the multiphoton excitation when single-cycle pulses are used.
The tunable values of CEP also serve as an additional control knob
in that process.
The control of the population transfer to the excited state of the
system (i.e. degree of multiphoton absorption) is demonstrated first
on the simplified two-level model and then applied to the atomic
caesium (Cs).

Although the ultrashort duration of the single-cycle pulses suggests
the inspection of the associated phenomena in the time-domain, in
many cases, especially in the range of weak- to intermediate- pulse
intensities \cite{silberberg_2ph_1plus1,dantus_exp_review1_2,
silberberg_2ph_nonres1_2, silberberg_antiStokes_Raman_spect,
leone_res_nonres_raman_control, gersh_murnane_kapteyn_Raman_spect,
amitay_2ph_inter_field1}, frequency-domain analysis provides is
preferable.
As frequency domain is a domain where the actual pulse shaping is
applied, it is also the domain of the powerful rational pulse
design, which is based on the identification of the photoinduced
multiphoton pathways, once the complete photo-excitation picture of
the system is available.
The typical bandwidth of the pulses used in the recent control
experiments is 30-50nm (fwhm), corresponding to the shortest
possible pulse duration of 20-30fs (transform-limited pulse).
The spectral components available within this bandwidth contribute
to the pathways with the single possible combination of the number
of absorbed and emitted photons.
For example for the two-photon absorption process, such combination
is the sequential absorption of two photons, while for the Raman
process it is absorption of one photon and emission of one photon.
Consequently, all the corresponding two-photon pathways in both
cases have the same global phase, associated with CEP, and their
interference does not depend on its specific value.

Recently, the shaped pulses with octave spanning bandwidth became
available \cite{dantus_octave_shaping}.
The shaped single-cycle pulses are expected to emerge in the nearest
future.
Their ultrabroad bandwidth provides additional types of pathways
between the ground and the excited states of the system to be used
for the control over the multiphoton absorption.
As such, there are several possible combinations of the number of
absorbed and emitted photons contributing to the different types of
pathways.
However, as it is shown below, the full control over their
interference is only possible for the pulses with stabilized CEP.
For example, consider the excitation of the system from the ground
state $\left|g\right>$ of energy $E_{g}$ to the excited state
$\left|f\right>$ of energy $E_{f}$ which can be accessed with the
absorption of two photons with frequencies that sum up to the
two-photon resonance $\Omega_{fg}=(E_{f}-E_{g})/\hbar$:
$\Omega_{fg}=\omega_{seq}+\omega_{seq}'$, where $\Omega_{fg} >
\omega_{seq}, \omega_{seq}'$.
In addition, the octave-spanning spectrum provides the Raman-type
pathways \cite{silberberg_antiStokes_Raman_spect,
leone_res_nonres_raman_control, gersh_murnane_kapteyn_Raman_spect}
that lead to the same excited state with absorption of one and
emission  of one photon. The frequencies of the photons sums up to
the two-photon transition $\Omega_{fg}=\omega_{Ram}-\omega_{Ram}'$,
where $\omega_{Ram} > \Omega_{fg} > \omega_{Ram}'$.
The ground and the excited states of the system are coupled via the
manifold of the intermediate states $\left|n\right>$ that for the
case of non-resonant two-photon absorption are not accessible by the
one-photon process.

The total amplitude of the excited state $A^{(2)}_{f}$ is
proportional to the Fourier transform of the square electric field
of the laser pulse
\begin{equation}\label{eq-1}
    A^{(2)}_{f}\propto\int_{-\infty}^{\infty}
    E^{2}(t)\textrm{exp}[i\Omega_{f} t] dt,
\end{equation} %
where
\begin{equation}\label{eq-2}
    E(t)=\frac{1}{2}{\cal{E}}(t)\textrm{exp}\left[i\left(\omega_{0}t+\phi_{CE}\right)\right]+C.C.,
\end{equation}
${\cal{E}}(t)$ - is gaussian temporal complex amplitude of the
pulse, $\omega_{0}$ - is the carrier frequency,  $\phi_{CE}$ - is
the CEP, and $C.C.$ stands for the complex conjugate term.
The electric field of the pulse $E(t)$ corresponds to the spectral
field $E(\omega)={\cal{E}}(\omega)\textrm{exp}[i\phi_{CE}]+C.C.$, %
where ${\cal{E}}(\omega)$ - is the spectral complex envelope of the
pulse
${\cal{E}}(\omega)=\left|{\cal{E}}(\omega)\right|\textrm{exp}[i\Phi(\omega)]$,
$\left|{\cal{E}}(\omega)\right|$ - is the spectral amplitude, and
$\Phi(\omega)$ is the relative spectral phase.
The shortest possible pulse for a given spectrum, referred as
transform-limited (TL) pulse, corresponds to $\Phi(\omega)\equiv$0
for any $\omega$.
The global phase $\phi_{CE}$ modifies the exact temporal shape of
the pulse, but not affects its duration.
The excited state population $P_{f}=\left|A^{(2)}_{f}\right|^2$ is
expressed in the frequency domain as a summation over the amplitudes
contributed by the sequential $A^{(2)}_{S}$ and the Raman
$A^{(2)}_{R}$ types of pathways
\begin{equation}\label{eq-3}
    P_f  = \left| {\mu _{seq}^2 A^{(2)}_{S} (\Omega _{f} ) + \mu _{Ram}^2
    A^{(2)}_{R} (\Omega_{f} )} \right|^2,
\end{equation}
where $\mu _{seq}^2$ and $\mu _{Ram}^2$ are effective non-resonant
two-photon sequential and Raman transitional dipole moments
correspondingly, and
\begin{eqnarray}\label{eq-4}
    A^{(2)}_{S} (\Omega) &=& \textrm{exp}[i2\phi_{CE}] \int^{\infty}_{0}
    {\cal{E}}(\omega){\cal{E}}(\Omega-\omega) d\omega \\ \nonumber
    A^{(2)}_{R} (\Omega) &=& 2\int_{\Omega}^{\infty}
    {\cal{E}}(\omega){\cal{E}}^{\ast}(\omega-\Omega) d\omega
\end{eqnarray}
For the pulse with stabilized CEP equals $\phi_{CE}$ these two types
of pathways have relative phase of $\Delta\phi=2\phi_{CE}$, which
dictates the nature of the interference between them.

\begin{figure} 
$$\includegraphics[width=8cm]{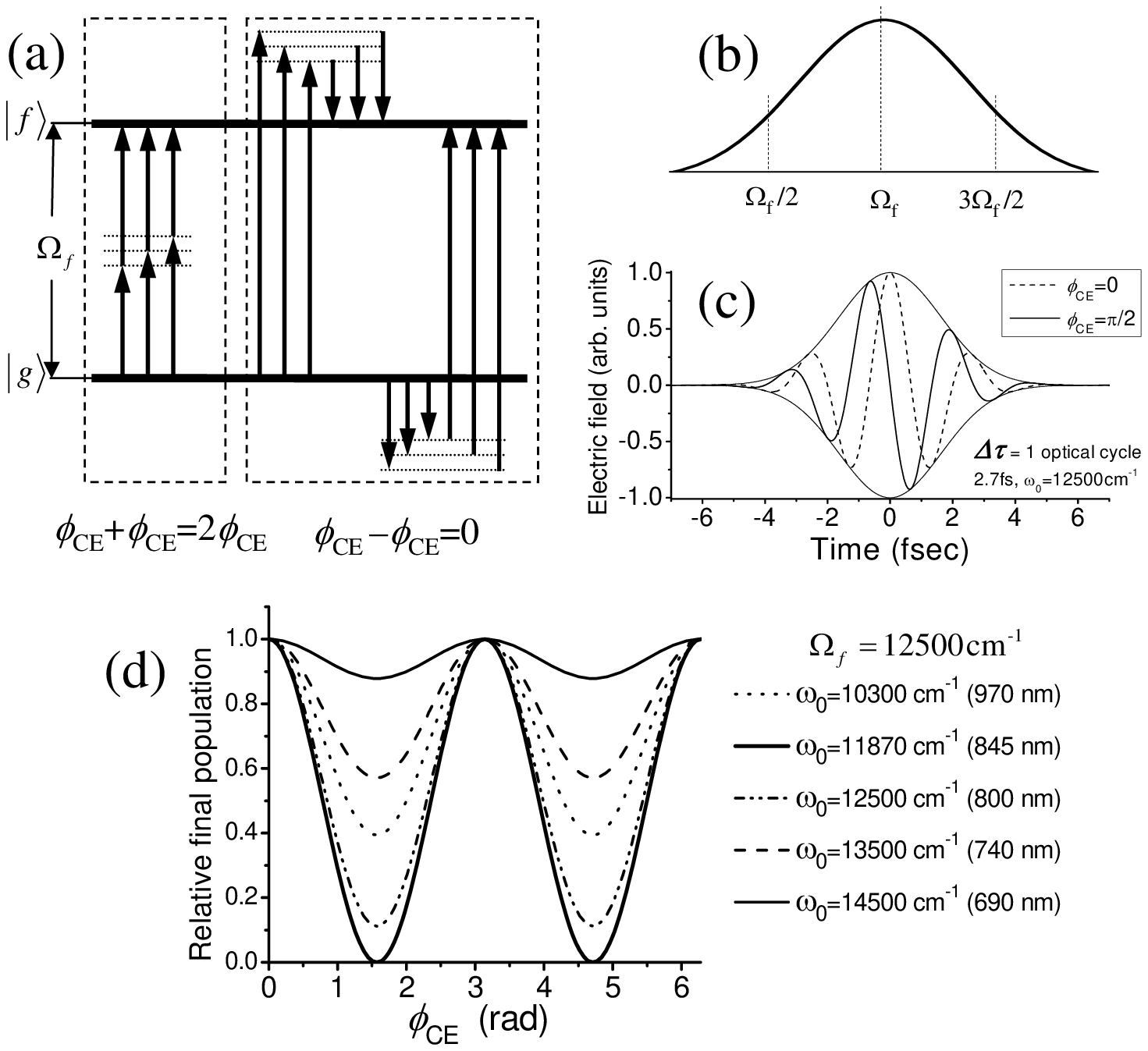}$$     
\caption{
(a) Two-level system excitation scheme. The two types of pathways --
the sequential and the Raman pathways are indicated.
(b) Ultrabroad spectrum of the single-cycle pulse. The spectrum is
centered at $\Omega_{fg}$ to provide $\Omega_{fg}/2$ photons for the
sequential pathways as well as $3\Omega_{fg}/2$ photons for the
Raman pathways.
(c) The temporal electric field of the single-cycle pulse with
different CEP.
(d) Degree of the two-photon absorption for the pulses with
different $\phi_{CE}$ and different carrier frequency $\omega_{0}$,
normalized to the case of $\phi_{CE}=0$.
%
%
} \label{fig_1}
\end{figure}

Consider the model quantum system with two-photon transition of
$\Omega_{f}$=12500cm$^{-1}$.
The sequential and Raman types of pathways are shown schematically
on Figure~\ref{fig_1}(a).
The duration of the pulse with fwhm equals to one optical cycle of
its carrier frequency of $\omega_{0}=12500$cm$^{-1}$ (800nm) is
2.7fs, and the corresponding spectral width of such pulse is
5425cm$^{-1}$.
The typical spectrum of the pulse and its temporal electric field
are drawn on Figure~\ref{fig_1}(b) and (c), respectively.
As opposed to the case of the two-photon absorption with pulses of
narrower spectrum, where the considered optimal spectrum is centered
at the half of the two-photon transition $\omega_{0}=\Omega_{f}/2$,
in the case of single-cycle pulse in order to utilize the
interference between different types of pathways the spectrum of the
pulse should be centered near $\Omega_{f}$.
This spectral positions ensures that the pairs of photons around
$\Omega_{f}/2$ contributing to the sequential type of pathways, as
well as the pairs of photons consisted of one $3\Omega_{f}/2$-photon
and one $\Omega_{f}/2$-photon contributing to Raman type of pathways
are provided.
The relative weight of each type of pathways is set on one hand by
the corresponding values of the effective two-photon transitional
dipole moments $\mu^{2}_{seq}$ and  $\mu^{2}_{Ram}$, and on the
other hand by the spectral amplitude profile of the pulse.
In our model system we set $\mu^{2}_{seq}=\mu^{2}_{Ram}$ for
simplicity.

We examine now the dependence of the transition probability to the
excited state on the value of $\phi_{CE}$.
The electric field of the pulses with different $\phi_{CE}$ is shown
on Figure~\ref{fig_1}(b).
In the frequency domain these pulses correspond to different global
spectral phase.
The beating of the excited state population against the $\phi_{CE}$
is traced on Figure~\ref{fig_1}(d) for different $\omega_{0}$.
Each trace is normalized to the case of  $\phi_{CE}=0$, where the
maximum of the pulse electric field oscillations coincides with the
maximum of the envelope.
The amplitudes $A^{(2)}_{S}$ and $A^{(2)}_{R}$ are in phase in this
case and interfere constructively.
Their relative weight depends on the spectral position, and
determines the maximum of the modulation depth $M =2\frac{ max
\left(P_{f}\right)-min \left(P_{f}\right)} {max
\left(P_{f}\right)+min \left(P_{f}\right)}$.
Therefore, we have $M_{\phi_{CE}=0}$=0.
As it is shown on Figure~\ref{fig_1}(d), the maximal values of $M$
are obtained for the case of $\phi_{CE}=\pi/2$.
When the spectrum is shifted to the higher frequencies, the relative
weight of the Raman pathways dominates the sequential ones, while
when the spectrum is shifted to the lower frequencies -- sequential
pathways dominate.
The higher is the difference in the relative amplitudes, the lower
is the modulation depth $M$.
The maximal contrast in the modulation depth $M_{\phi_{CE}=\pi/2}$=2
was achieved for $\omega_{0}=11870$cm$^{-1}$ [Figure~\ref{fig_1}(d),
thick solid line], corresponding to the red shift of the spectrum as
compared to the two-photon resonance $\Omega_{f}=12500$cm$^{-1}$.

\begin{figure} 
\includegraphics[width=8cm]{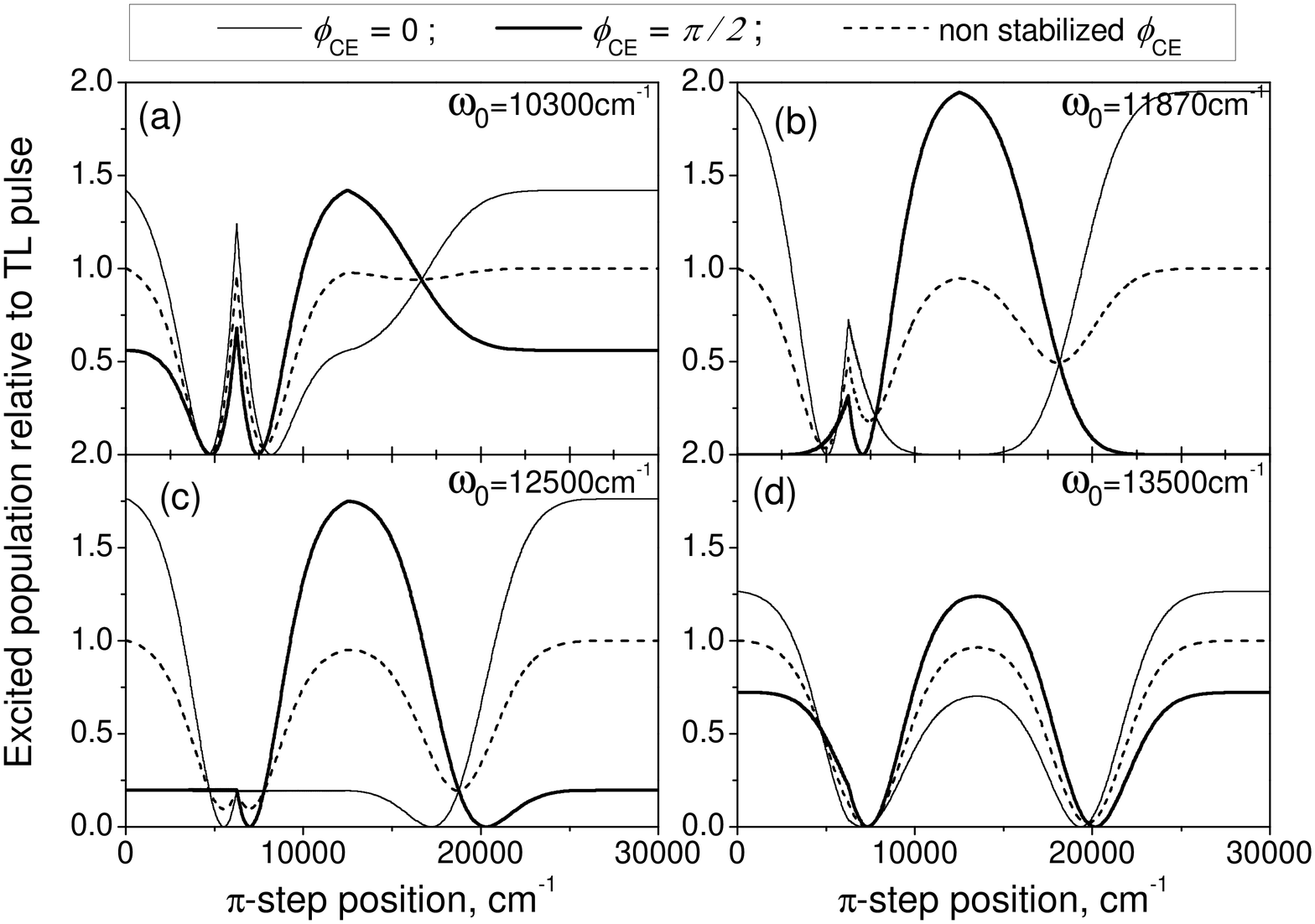}     
\caption{
Coherent control of two-photon absorption in the two-level model
system.
Results for a $\pi$-step phase patterns with different
$\omega_{step}$ are shown for single-cycle pulses of different
carrier frequencies $\omega_{0}$ and different $\phi_{CE}$.
The two-photon absorption is normalized to the values of TL pulse
with non-stabilized $\phi_{CE}$.
} \label{fig_2}
\end{figure}

Next, as a test-case for the coherent control of two-photon
absorption, we consider the phase-shaped pulses, having a $\pi$-step
spectral phase pattern, while the step position $\omega_{step}$ is
scanned along the pulse spectrum.
The control of the excited state population is examined for the
pulses with different positions of the spectrum [different
$\omega_{0}$], and the results are plotted on Figure~\ref{fig_2}.
For each spectrum we show the traces of the excited state population
as a function of $\omega_{step}$ for $\phi_{CE}=0$,
$\phi_{CE}=\pi/2$, and the trace averaged over all possible values
of $\phi_{CE}$.
The latter corresponds to the case of the not stabilized CEP.
The calculated results are normalized to the population excited by
the unshaped pulse with $\phi_{CE}$=0.
The results strongly emphasize the role of the CEP stabilization of
the few-cycle pulses and the dependence of the excitation yield on
the actual value of $\phi_{CE}$.
The case of $\omega_{0}=11870$cm$^{-1}$ is shown on panel (b).
For the pulse with $\omega_{step}$ at the very low frequencies the
excited population is the same as shown on Figure~\ref{fig_1}(d).
When $\omega_{step}$ is set at the high frequency region, all the
spectrum experience additional constant phase of $\pi$.
Here the global phase of the shaped pulses has been changed from
$\phi_{CE}=0$ and $\phi_{CE}=\pi/2$ to $\phi_{CE}=\pi$ and
$\phi_{CE}=3\pi/2$, however, the excitation yields remain the same
as can be expected from Figure~\ref{fig_1}(d).

The picture is different when the $\omega_{step}$ is set in the
middle of the spectrum.
Take for example $\omega_{step}=12500$cm$^{-1}$ with $\phi_{CE}=0$.
The spectral components near $\Omega_{f}/2$ of the sequential
two-photon pathways have zero relative phase and interfere
constructively within $A^{(2)}_{S}(\Omega_{f})$.
The spectral components of the Raman pathways has different phases
[0 for $\omega<\omega_{step}$ and $\pi$ for
$\omega\ge\omega_{step}$] as they are coming from the different
spectral regions.
Consequently, the Raman pathways interfere constructively within
$A^{(2)}_{R}(\Omega_{f})$, but there is a global phase of $\pi$
between the amplitudes $A^{(2)}_{S}(\Omega_{f})$ and
$A^{(2)}_{R}(\Omega_{f})$.
Hence, while for the unshaped pulse $\phi_{CE}=0$ corresponds to the
completely constructive interference between
$A^{(2)}_{S}(\Omega_{f})$ and $A^{(2)}_{R}(\Omega_{f})$, for the
shaped pulse with $\omega_{step}=12500$cm$^{-1}$ the interference is
completely destructive and the overall excited state population is
zero.
When the similar analysis is applied to the case of
$\phi_{CE}=\pi/2$, we obtain that while for the unshaped pulse the
excited population is zero due to the destructive intergroup
interference, it is turned into the constructive one by the
corresponding pulse shaping with $\omega_{step}=12500$cm$^{-1}$.

\begin{figure} 
\includegraphics[width=8cm]{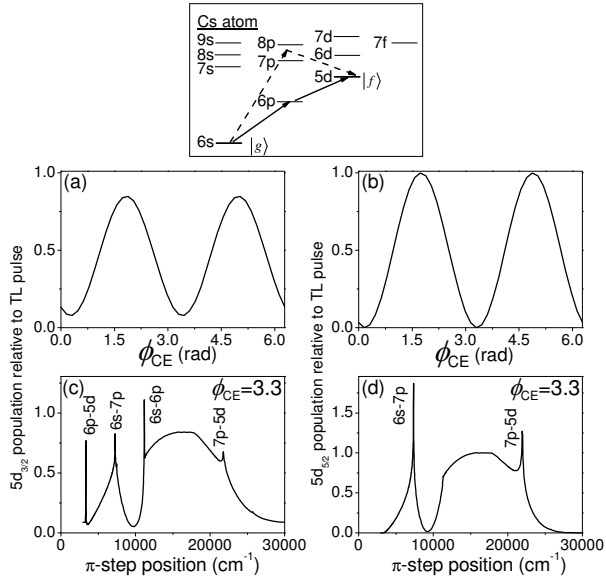}     
\caption{
Upper panel: Schematic diagram of the Cs atom (fine structure
splitting is not shown).
The ground state is $\left|g\right>\equiv$6s, the two excited states
are $\left|f_{1}\right>\equiv$5d$_{3/2}$ and
$\left|f_{2}\right>\equiv$5d$_{5/2}$ .
The examples of the sequential and Raman pathways are shown by solid
and dashed lines respectively.
Lower panel: %
Population of the excited states of Cs, calculated for single-cycle
[fwhm=2fs] pulses with $\omega_{0}$=16667cm$^{-1}$. The results are
normalized to the maximal population of the 5d$_{5/2}$ state.
(a),(b) The population for different values of $\phi_{CE}$.
(c),(d) The population excited with the phase-shaped pulses for
$\phi_{CE}$=3.3. The pulses are shaped with $\pi$-step phase
patterns at different $\omega_{step}$. The enhanced transitions
associated with the peaks are indicated. } \label{fig_3}
\end{figure}

Our two-state frequency domain perturbative model is verified here
by the exact solution of the time-dependent Schr\"{o}dinger equation
for the atomic Cs irradiated by the shaped single-cycle pulses.
We used numerical propagation by fourth-order Runge-Kutta method.
The 6s state was considered as the ground state, while the
5d$_{3/2}$ and $5d_{5/2}$ states -- as the excited states.
The excitation scheme is shown on Figure~\ref{fig_3}: the sequential
pathways are drawn by the solid line, the Raman pathways -- by the
dashed line.
The two-photon coupling between the ground and excited states is
provided by the manifold of the p-states of the atom.
Overall, we have considered 6s-9s, 6p-8p, and 5d-7d states,
including also their fine structure splitting.
The information on the atomic levels and the corresponding
transitional dipole moments can be found in \cite{NIST,
safronova_Cs_12, blundell_Cs, safronova_Cs_3}.
As opposed to our two-state model, in real atom the values of the
effective two-photon transition dipole moments are generally not
equal in magnitude, and also can have different sign, depending on
the specific atomic structure.
Here we first examine the dependence of the final states population
on the $\phi_{CE}$ value of the excitation pulse, and then
demonstrate the results for the shaped pulses.

The energies of the 5d$_{3/2}$ and 5d$_{5/2}$ states are
$\Omega_{f_{1}}$=14499.3cm$^{-1}$ and
$\Omega_{f_{2}}$=14596.8cm$^{-1}$, respectively.
We have found that control of the excited population is most
efficient when the spectrum of the pulse is centered at
$\omega_{0}$=16667cm$^{-1}$ [600nm].
The  optical cycle corresponding to this wavelength has duration of
2fs, and the bandwidth associated with the single-cycle pulse is
7350cm$^{-1}$.
The calculated results are shown on Figure~\ref{fig_3}(a),(b).
The maximal modulation value $M$=2 for the 5d$_{5/2}$-state
population is obtained with $\phi_{CE}$=3.3rad.
For the 5d$_{3/2}$ state the maximal modulation value obtained is
$M$=1.7 corresponding to $\phi_{CE}=3.3$.

On Figure~\ref{fig_3}(c),(d) we demonstrate results for the
excitation of Cs atoms for the case of $\phi_{CE}$=3.3 for various
$\pi$-step positions.
When $\omega_{step}=\omega_{0}$, as it was expected from the
analysis of the two-level system, the destructive interference
between the sequential and the Raman pathways is turned into the
constructive one, and the two-photon absorption for such shaped
pulses is maximal for a given spectrum.
The two-photon excitation process of Cs with broadband spectrum at
$\omega_{0}$=16667cm$^{-1}$ has resonance-mediated nature
\cite{silberberg_2ph_1plus1}.
It is reflected in the response to the excitation with $\pi$-step
phase shaped pulses.
When the $\omega_{step}$ is set exactly at the  frequency
corresponding to initial-to-intermediate or intermediate-to-final
state resonance transition, there is a significant enhancement in
the two-photon absorption.
The mechanism of such enhancement is conversion of destructive
interference between the sequential two-photon pathways associated
with the TL pulse into constructive one by the appropriate pulse
shaping \cite{silberberg_2ph_1plus1}.
Consequently, each peak in the two-photon absorption that appear on
Figure~\ref{fig_3}(c),(d) correspond to the specific resonance of
the system.

To summarize, we have demonstrated coherent control of the
two-photon absorption in the single-cycle regime.
The important role of the CEP in control over the interference
between the sequential and Raman-type pathways is revealed in the
frequency domain analysis applied to the two-level model system,
extended to atomic Cs, and verified by the exact solution of the
time-dependent Sr\"{o}dinger equation.
The results for coherent control of two-photon absorption are
compared with the case of non-stabilized CEP.


\end{document}